# ARTICLE

# Microfluidic bioanalytical flow cells for biofilm studies: A review

Mohammad Pousti,[a] MirPouyan Zarabadi,[a] Mehran Abbaszadeh Amirdehi,[a] François Paquet-Mercier,[a] and Jesse Greener*[a,b]

Bacterial biofilms are among the oldest and most prevalent multicellular life forms on Earth and are increasingly relevant in research areas related to industrial fouling, medicine and biotechnology. The main hurdles to obtaining definitive experimental results include time-varying biofilm properties, structural and chemical heterogeneity, and especially their strong sensitivity to environmental cues. Therefore, in addition to judicious choice of measurement tools, a well-designed biofilm study requires strict control over experimental conditions, more so than most chemical studies. Due to excellent control over a host of physiochemical parameters, microfluidic flow cells have become indispensable in microbiological studies. Not surprisingly, the number of lab-on-chip studies focusing on biofilms and other microbiological systems with expanded analytical capabilities has expanded rapidly in the past decade. In this paper, we comprehensively review the current state of microfluidic bioanalytical research applied to bacterial biofilms and offer a perspective on new approaches that are expected to drive continued advances in this field.

## Introduction

Living biofilms are one of the most challenging materials to study quantitatively. Analytical tools based in electrochemistry, spectroscopy and quantitative imaging are beneficial, but their full potential can only be realised with proper control over the liquid phase. Therefore, microfluidics is quickly becoming a fundamental aspect of modern bioanalytical sciences.

### Literature review

We rigorously analysed all published papers that used microfluidic devices to study bacterial biofilms. We categorised the applied measurement modalities to assemble a global overview of the state of microfluidic bioanalytical methods. Therefore, this work complements and goes beyond previous reviews of the use of microfluidics which have previously focused on certain specific phenomena such as biofilm formation, adhesion and detachment;[1-3] biofilm susceptibility to antibiotics and toxicity testing;[4] and electroactive biofilms including those used for microbial fuel cells.[5-8]

Figure 1a shows the annual numbers of manuscripts published since 2002 that report microfluidic studies of bacterial biofilms. These numbers grew rapidly between 2007 and 2016, with nearly 300 papers published by mid-2018. It is difficult to know if the apparent levelling off after 2016 is significant or if it is just an aberration related to the estimated number of papers in 2018 based, which was based on an extrapolation to the end of the calendar year. We note that the graphic is not comprehensive across all microfluidic domains because it does not include alternative platforms such as digital[9] and paper[10] microfluidics. Also excluded are "static" microscale devices, which can share microfabrication techniques and materials with channel-based microfluidics but do not involve flow. Keywords from the titles of all analysed papers are presented in graphical format in Figure 1b. This graphic shows some of the major drivers of the field, based on title keywords. A text file listing the most popular title keyword is given in the ESI. Based on a detailed reading of nearly 300 relevant papers since 2002, a global breakdown of characterisation methods is given in Figure 1c. Here multiple uses of different techniques within the same category were only counted once. For example, work using multiple spectroscopic or electrochemical measurements was considered a single instance in either category. The categories identified in this figure define the subject matter and the organization of this paper.

Optical microscopy is the historically dominant measurement modality. This includes transmission modes, which enable quantification of biofilm mass and dynamic motion via optical density and time-lapse imaging, respectively.[11] Epifluorescence methods provide high-contrast imaging and can target specific biological and biochemical groups via fluorescent probe molecules. High-resolution microscopy, which we define here as either providing higher spatial resolution or dimensionality than transmission and fluorescence microscopy, has been the second most frequently used in published microfluidic biofilm studies. This category includes confocal laser scanning microscopy (CLSM), atomic force microscopy (AFM) and electron microscopy such as scanning and transmission electron microscopy (SEM, TEM). With few exceptions, AFM and

[a] Département de chimie, Faculté des sciences et de génie, Université Laval, Québec City, Québec G1V 0A6, Canada.
[b] CHU de Quebec Research Centre, Laval University, 10 rue de l'Espinay, Quebec City (QC) G1L 3L5, Canada.
† Footnotes relating to the title and/or authors should appear here.
Electronic Supplementary Information (ESI) available: [details of any supplementary information available should be included here]. See DOI: 10.1039/x0xx00000x

ARTICLE

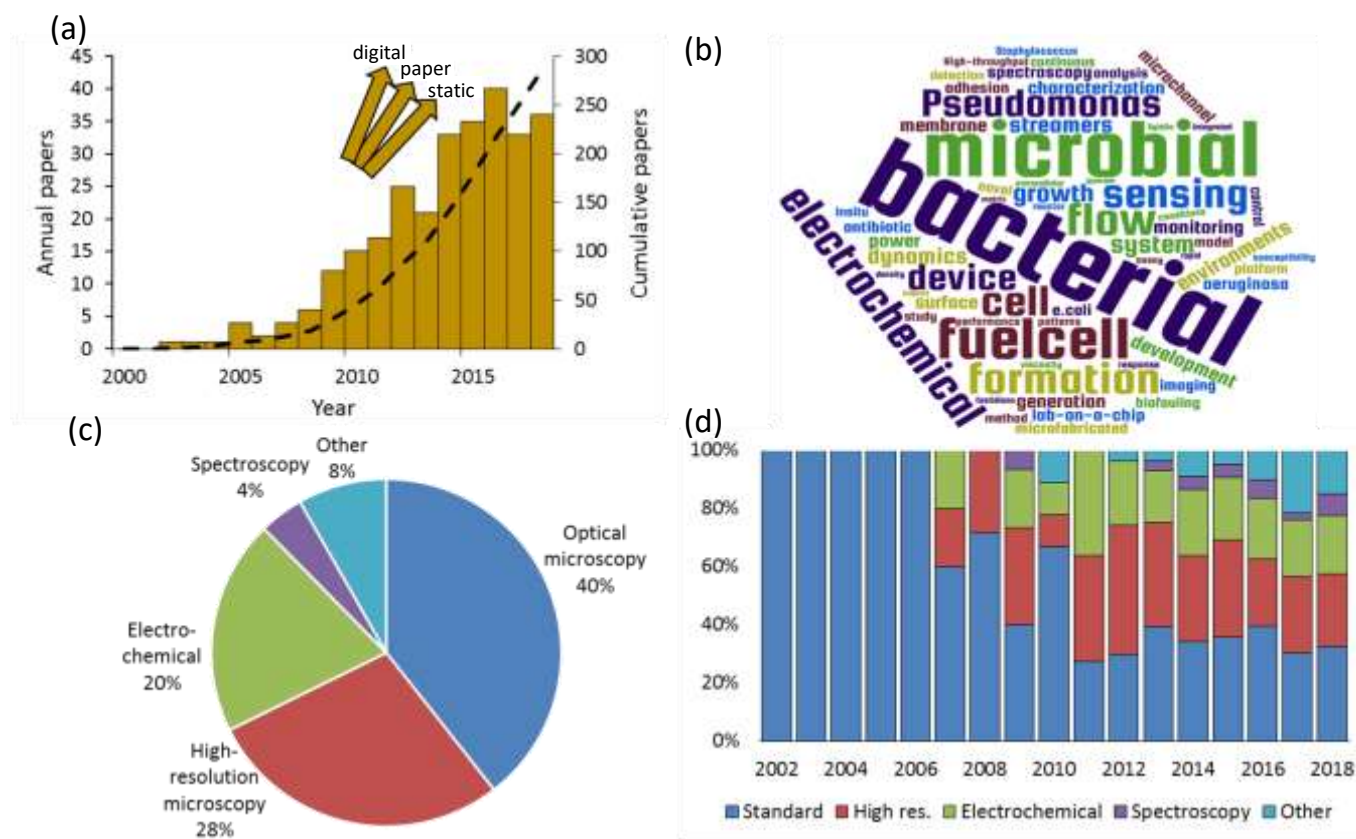

Figure 1. Evaluation of the literature on microfluidic biofilm studies. (a) Annual and cumulative number of publications using microfluidic studies of bacterial biofilms (gold bars and black dashed line, respectively). Contributions do not include digital, paper or static microfabricated environments. The year 2018 is adjusted based on extrapolated number of papers to the end of the calendar year. (b) Word cloud featuring the most popular keywords from all article titles. Break-down of techniques used in journal articles since 2002 (c) and on a year-by-year basis (d).

electron microscopy are only applicable to *ex situ* microfluidic biofilm studies. On the other hand CLSM is easily applied *in situ*. This almost certainly accounts for the majority use of CLSM in this category (> 60%). By far, the most important non-imaging approaches are electrochemical. This has been driven by a growing interest in electroactive biofilms, though some electrochemical techniques can even be applied to non-electroactive biofilms, further increasing their relevance. Spectroscopic studies represent an important undercurrent in the literature. This category is mainly composed of vibrational techniques (Raman and infrared spectroscopy). Finally, other techniques and emerging approaches are discussed. Here we include important developments and trends toward on-chip implementation of specialised microscope imaging techniques, electrochemistry and spectroscopy, such as on-chip magnetic resonance imaging, as well as emerging approaches of mass spectrometry, respirometry. These methods are also compared in terms of their ability to meet a range of characterization requirements for biofilm studies.

True to the concept of lab-on-a-chip, fully one third of the published studies applied more than one characterisation category in a single study. This also supports recent conclusions that multiple analytical techniques applied to the native biofilm are required to obtain most meaningful results and continued advancement of the field.[12] We also note in passing that approximately 15% of all relevant papers used numerical simulations to predict the flow properties within the microchannels. The final global characteristic extracted from this literature search was the significant diversification of the analytical modalities used over time (Figure 1d). Although microscopy was dominant in early biofilm work, a decisive shift has occurred toward other approaches. This observation matches the trends in non-microfluidic experiments, which increasingly involve chemical and bio-chemical characterisation approaches such as respirometry,[13] spectral microscopy and chemical imaging[14] and electrochemistry.[15] Taken together, these trends indicate that biofilm research is becoming a driver for the development of integrated multifunctional microfluidic technology. To appreciate the need for such sophistication, a short summary of bacterial biofilms is presented next.

**Bacterial biofilms, grand analytical challenges**
Sessile microbes, i.e., those living at surfaces, preferentially form biofilms. Contrary to their name, biofilms are three-dimensional materials with heterogeneous structural and chemical properties. In addition to living bacteria, biofilms are usually composed of an extracellular polymeric substance (EPS) consisting of biomacromolecules such as proteins, polysaccharides and nucleic acids. The EPS helps to regulate molecular diffusion, microbe-to-microbe interactions and other biochemical functions, while simultaneously maintaining the physical and mechanical properties necessary for survival under hydrodynamic conditions.[16,17] Therefore, microbial biofilms can form their own niche to achieve homeostasis in harsh or fluctuating conditions. This special property allowed microbial biofilms to become one of the first multicellular life forms on ancient Earth, approximately 3 to 4 billion years ago. Some of these ancient biofilms can still be found in certain niche environments today such caves[18] and in extreme temperature, acidic and anaerobic conditions[19-21] but are now joined by a diverse range of modern species that have proliferated to all corners of modern Earth.

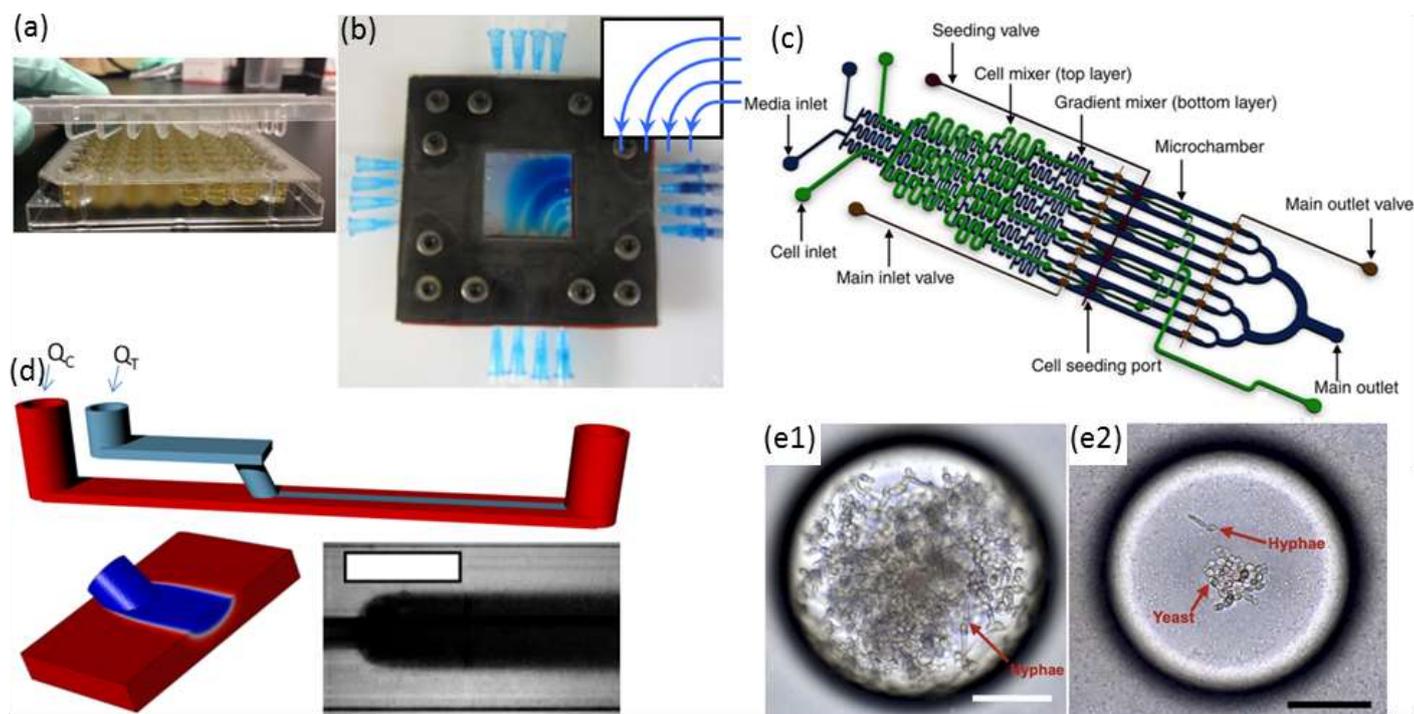

Figure 2. (a) Conventional Calgary device. (b) Microfluidic planar flow cell with height of 600 µm.[48] Inset shows the direction of flow. (c) Multi-level device featuring complex fluidic circuitry designed to generate concentration gradients with built-in mixers, purpose-specific inlets, valves and separated growth compartments.[64] (d) laminar flow-templating device for creating linear biofilms patterns using liquid-liquid interfaces to prevent growth in microchannel corners. (e) Encapsulated microenvironments containing yeast and hyphae forms of biofilm forming *Candida albicans* cells after encapsulation (e1) and the more robust yeast cells when incubated in a culture of *Pseudomonas aeruginosa* PA14.[69]

Recently, a rapid expansion in biofilm research is been driven by multiple areas of enquiry. In addition to fundamental questions focused on the relationship between biofilm structure and ecological function,[22] these drivers include applied research into marine and industrial fouling,[23,24] dental research,[25] indoor moulds,[26] virulent biofilms[27-30] and biocatalysis.[31,32] To complement the tremendous advances in molecular biology studies of biofilms in the last 10-15 years, greater attention has been placed on quantitative analysis under well-defined physiochemical conditions. In other words, modern bioanalytical investigations are increasingly directed toward improved control over growth conditions. This situation is especially true for biofilms, which efficiently respond to environmental cues by altering fundamental properties such as metabolic activity, biochemistry, physical structure and molecular mass transport.[14,33-35]

In the following section, we discuss different flow cell designs and the potential for microfluidic platforms to deliver enhanced control over key environmental parameters during biofilm studies.

## Control using microfluidics

Biofilm growth cells are either static or flow-based. Static reactors are attractive because they are straightforward in design and easy to use. For example, assays conducted in microwells are common, with fluid handling usually being conducted via manual or robotic pipetting. A major drawback is that planktonic bacteria and other suspended material can contaminate biofilms due to gravitational settling. This problem can be alleviated with modified growth cell designs such as the Calgary device (Figure 2a) in which biofilms are grown on pegs that are inserted into microwells from the top side. Periodic withdrawal of the biofilm-coated pegs for analysis leaves debris and foreign bacteria to settle at the bottom of the well.[36] Drawback include changes in the biochemical environment due to continuous nutrient depletion and by-product accumulation. Frequent sample handling also limits measurement reproducibility and can introduce contamination. In contrast, flow cells maintain a constant flux of nutrient solution against the biofilm in a sealed environment, which can significantly reduce all of these problems.

Moreover, flow is among the most important factors in determining fundamental biofilm behaviour due to its effect on imposed shear stress[37,38] molecular flux,[39] concentration gradients in and around the biofilm[40-43] and interactions with planktonic bacteria.[44] As such, flow affects most observable biofilm properties, including growth kinetics,[45] structure[46-49] and function.[50] However, due to the large liquid volumes required, standard flow cells present challenges in long-term studies over a comprehensive range of flow rates. Microfluidics can expand the benefits of standard flow cells while also offering other unique advantages. On-chip volumes at the µL scale or smaller require remarkably low flow rates, even during manipulation of hydrodynamic properties over a large dynamic range. Thus, nutrient solution consumption can be strongly limited, which is beneficial in long-term experiments. Unlike in miniature chemostats or millifluidic flow cells, dead volumes in microchannels are vanishingly small, ensuring predictable liquid/biofilm contact times and applied concentrations. Flexibility in channel design and fluid pumping sequences enables complex multi-step experimental sequences and rapid adjustment of physiochemical parameters via on-chip dilution[51,52] or integrated microvalves.[53] For example, a microfluidic planar flow cell facilitated the application of a wide range of concentrations to the biofilm culture environment (Figure 2b).[48] A range of rapid prototyping techniques exists for fabrication of different microchannel geometries,[54-56] enabling purpose-driven environments such as pseudo-porous and serpentine channels for the study of streamers,[57,58] highly parallelised channels for high-throughput assays,[59] complex fluidic circuitry[60] and other moving components.[61,62] For example, microfluidic concentration-gradient generators with complex circuitry for parallel biofilm assays using separate media and cell seeding inlets have been successfully applied in a number of studies (Figure 2c).[63,64] Studies on the effects of biofilms under moving three-phase interfaces can be efficiently conducted via the imposition of bubbles with well-defined sizes, velocities and wall interactions.[65-67] Nevertheless, biofilms are highly efficient at modifying their environment, potentially undermining the advantages above.[37,68] Solutions to these problems have been demonstrated, including templating biofilm growth using liquid-liquid (Figure 2d)[45] or liquid-gas[67] interfaces, confinement in microdroplets[68,69] (Figure 2e) and isolation methods with integrated heating elements[71] or via confinement by locally high shear forces.[72] Isolation of electroactive biofilms at the surface of an integrated electrode can also limit channel overgrowth.[73]

To build on the previous achievements summarised above, the challenge for future biofilm researchers consists of merging microfluidic biofilm flow cells with powerful *in situ* analytical tools. The resulting synergy is expected to enable future discoveries and drive continued growth of microfluidics applied to biofilms and bioanalytical chemistry generally.

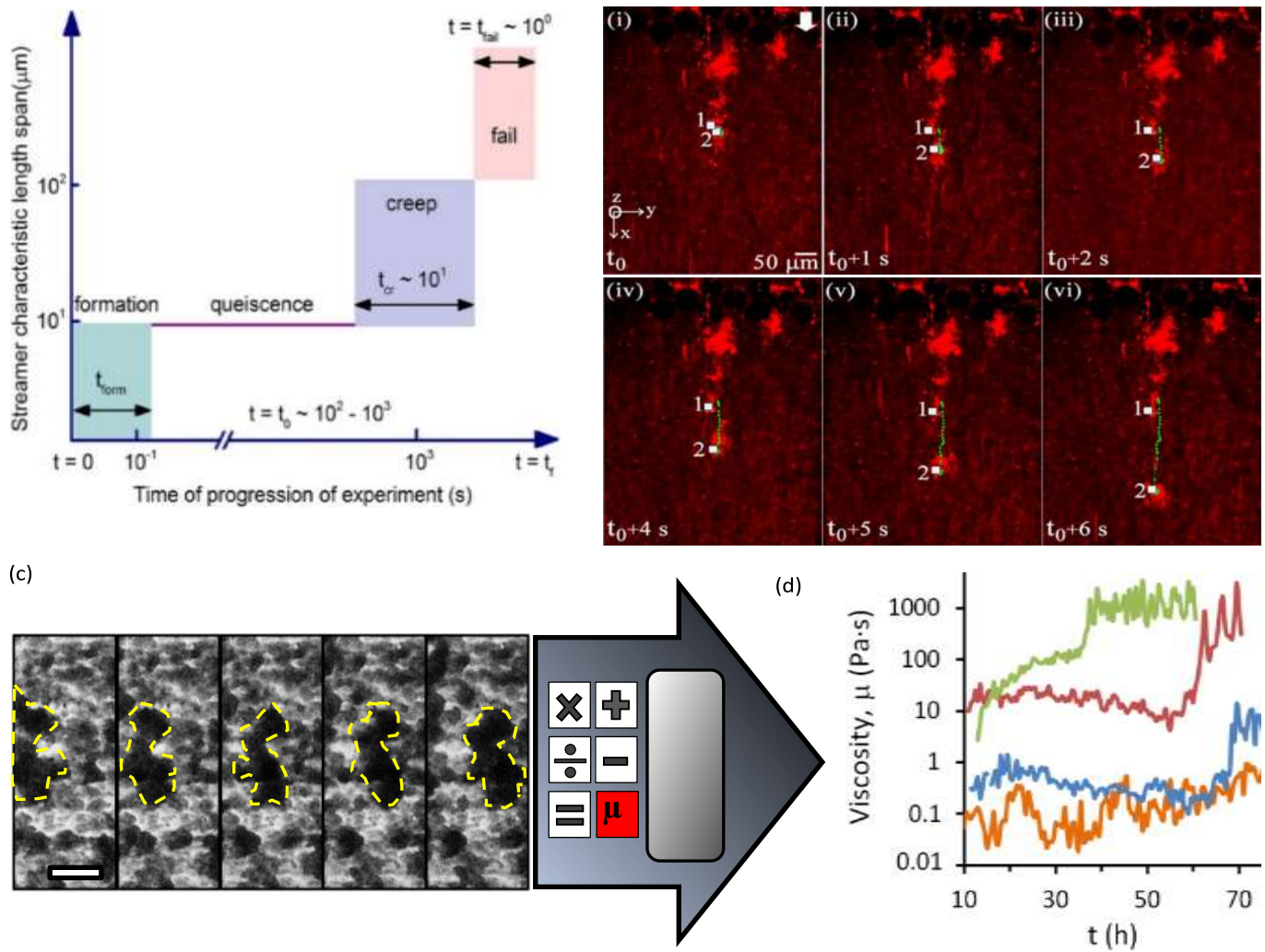

Figure 3. (a) Schematic of space-time scales of different regimes associated with biofilm streamer formation. (b) An example of streamer creeping motion occurring in the time segment shown in (a) immediately before failure.[91] (c) Moving biofilm segment at intervals of 1 h. (d) Resulting biofilm viscosity ($\mu_{biofilm}$) after application of a semi-empirical model for [NaCl] at 0 wt% (orange), 0.05 wt% (blue), 0.1 wt% (red) and 0.2 wt% (green).

## Microscopy

### Optical microscopy

Transmission microscopy and fluorescence microscopy were the most popular analytical tools for early microfluidic studies of bacterial biofilms. This is unsurprising because the combination of clear planar surfaces and the small internal dimensions of most microfluidic devices offer an excellent environment for undistorted, high-resolution imaging with no special requirements imposed on device fabrication or operation. In addition to supplying structural information at the micron to centimetre scale,[74] optical microscopy also enables quantitative optical density measurements.[11] Biofilm structural heterogeneity can also be quantified from microscope images via the coefficient of variance, fractal dimension and other image analysis techniques.[74,75] Because microscope-based analysis of biofilms has become well understood and is reviewed elsewhere,[1,33] we focus on certain alternative uses of optical microscopy before addressing other characterisation methods.

### Chemical tracers and time-lapse imaging

The main limitation of optical microscopy is that it is not directly applicable to chemical detection. Fluorescence microscopy can be transformed into a chemical imaging tool by taking advantage of the spectral properties of some fluorophores.[76] One microfluidic-compatible method loads freely diffusing fluorescent pH indicators into the biofilm from the liquid phase.[77-79] However, methods that avoid direct interaction between fluorescent probe molecules and the biofilm are preferable for reasons of biocompatibility and photostability. Recently, a pH-imaging microfluidic flow cell was reported that used a glass sealing layer with a layer of nanoparticles capable of metal-enhanced florescence using a pH sensitive fluorophore. Quantitative pH images of intact lactic acid biofilms (*Streptococcus salivarius*) were acquired with a fluorescence microscope while manipulating the hydrodynamic and chemical conditions for dental applications.[80] The continued development of such nanoparticles offers promising avenues for chemical imaging of other parameters, including ionic strength, electrolyte concentration and even potentially molecular activity markers within the biofilms.[81,82]

Optical microscopy can also be used to follow dynamic processes via time-lapse imaging with motion-tracking analyses. Such studies are readily conducted using microfluidics thanks to the ability to modulate or maintain the experimental conditions over both the short- and long-term. Typically, physiochemical processes that involve events on the order of seconds or minutes, include initial bacterial adhesion,[83] response to applied forces[61] and changing flow[84,85] or chemical conditions.[79] Slower processes are usually biological in nature, such as self-aggregation of pre-biofilm sessile cells,[86,87] growth kinetics,[45] changes to biofilm morphology and bacterial phenotype,[88-90] gene dynamics[91] and quorum sensing. Many of these phenomena require sensitivity over a wide range of time scales to characterise the entire process. For example, Figure 3a and 3b show biofilm stream lengths during different stages of *Pseudomonas fluorescens* development from adhesion to failure, with characteristic time scales ranging from less than 0.1 s to hours.[92] Another example is the use of a semi-automated tracking algorithm to monitor changes in viscosity of *Pseudomonas sp.* biofilms over a period of days under different nutrient concentrations[93] or ionic strengths.[94] In this work, the velocities of moving biofilm segments (e.g., Figure 3c) were fed into a semi-empirical numerical model to obtain the time-varying viscosity. Interestingly, after a certain time (which was correlated with the ionic strength), an abrupt thickening process was observed, resulting in increases in viscosity by more than an order of magnitude over a duration of less than ten hours (Figure 3d).[94] This observation enabled the identification of the rapid viscosity transition as the likely cause of streamer formation via a sudden partial detachment mechanism observed in another microfluidic experiment.[95]

Despite the use of chemical probes or motion tracking that expand the scope of optical microscopy using, fundamental constraints still remain. These include image resolution, dimensionality (two) and artefacts from biofilm segments outside of the focal plane. In the following section, we discuss high-resolution imaging techniques.

### High-resolution microscopy

We define high-resolution microscopy as imaging techniques providing either higher spatial resolution or dimensionality than transmission or fluorescence optical microscopy. Nanoscale resolution measurements on biofilm segments and individual biofilm components can be achieved using electron and scanning probe microscopy. The most prevalent methods include scanning electron microscopy (SEM) and atomic force microscopy (AFM). Due to the nature of these approaches and the sample preparation required, they typically cannot be conducted in closed microchannels. One notable study overcame this limitation for AFM by including specialised access ports through which AFM tips could be placed in contact with hydrated biofilms during growth.[96] This method could also be used in the future to introduce other scanning probes into microchannels, such as the powerful emerging technique of scanning electrochemical microscopy.[97] As well, new developments in self-contained microfluidic vacuum devices may open the way for the application of SEM for *in situ* measurements.[98] Because nanoscale resolution is not required for most biofilm studies, no strong push has been made to overcome the barriers outlined above.

The arrival of widely available confocal laser scanning microscopy (CLSM) has been credited as an important factor in the re-intensification of biofilm research in the 1990s. In contrast to regular fluorescence microscopy, CLSM excites fluorophores using a scanning laser in the x-y plane while z-axis resolution is achieved with a combination of stage positioning and confocal optics. Together with image reconstruction software, the acquisition of so-called image stacks enables measurements of biofilm heterogeneity in all three dimensions, yielding more representative and information-intensive images.[99] Similar to other optical microscopy techniques, CLSM is well-suited to most microfluidic environments and can accommodate chemically sensitive fluorescent probes and motion-tracking analysis, as discussed above. Although CLSM is fundamentally diffraction-limited it is included among the high-resolution techniques because the focused laser source and the confocal aperture enable practical enhancements to resolution and image dimensionality compared to most optical microscopy methods. As voxels are scanned one-by-one, the addition of a grating and array detector enables emission-side spectral imaging as a standard option for most CLSM instruments.

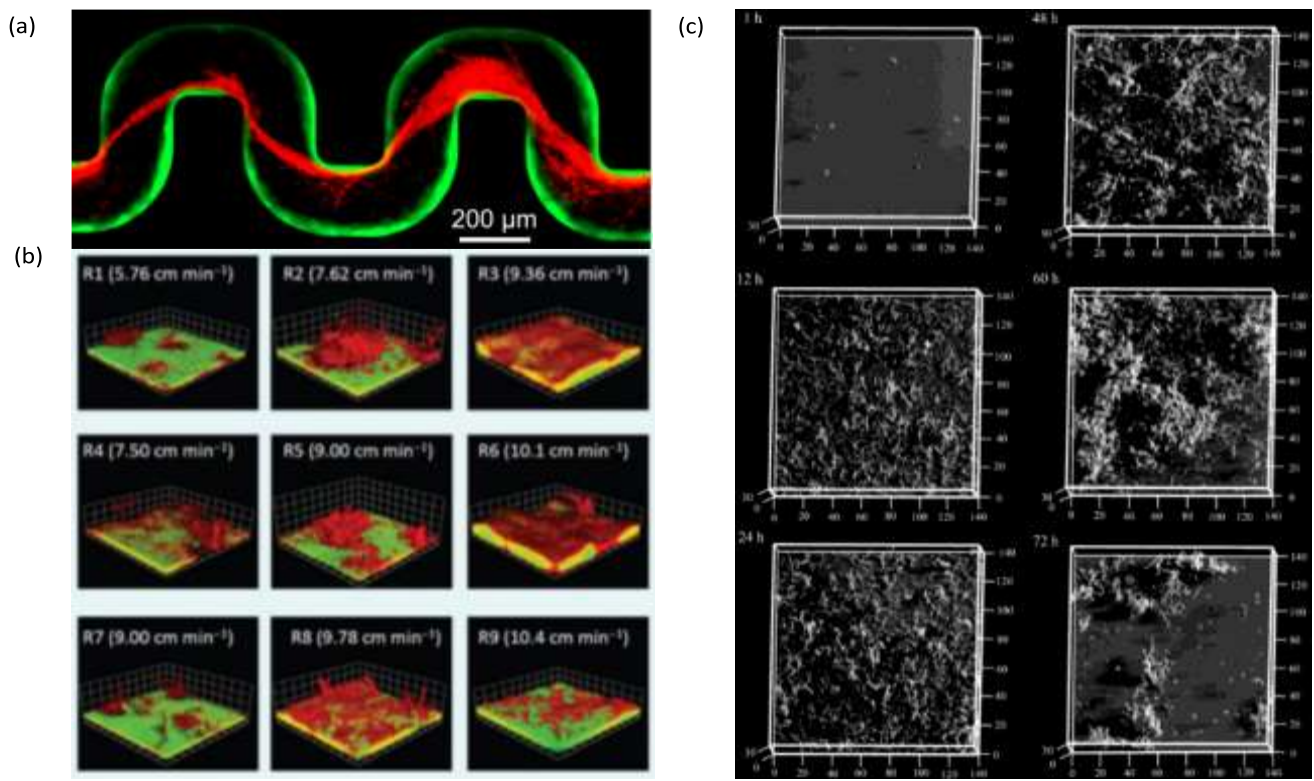

Figure 4. (a) Captured *Pseudomonas aeruginosa* bacteria (red) by pre-formed streamers in serpentine channels from a biofilm of the same type from previous inoculation and growth (green). [68] (b) Dual-species 7-day-old biofilms of *Pseudomonas aeruginosa* (green) and *Flavobacterium sp.* (red) under a range of local flow rates showing the representative biofilm morphology at each of the nine regions in a microfluidic flow velocity gradient system. [105] (c) Confocal reflection microscopy images of *S. mutans* NBRC13955 biofilm growth in a microfluidic device at different growth times. [107]

A useful application of CLSM is the study of streamers, which have notably distinctive three-dimensional patterns and can form efficiently due to laminar secondary flow streams near microstructures[2] or at channel turning points.[68] For example, with the use of bacteria carrying different fluorophores, CLSM enabled the study of the trapping kinetics of planktonic bacteria via pre-established streamers in a serpentine microchannel (Figure 4a). Another application combined time-lapse CLSM imaging at the attachment surface with electrochemical and vibrational spectroscopy to confirm a restructuring process in *Pseudomonas fluorescens* biofilms which was responsible for the formation of mushroom structures at the early growth stages.[100] Other reports include the use of CLSM to assess the effect of shear forces on adhesion, structure, and planktonic cell proliferation,[101] the effect of Reynolds number on bacteria distribution within biofilms[47] and high-throughput assessment of antimicrobials.[102] CLSM is also beneficial for examination of multi-species biofilms due to its high spatial resolution and ability to identify different bacterial lines expressing different fluorescent proteins. This aspect is regularly exploited in microfluidic studies to promote multispecies biofilm studies under controlled growth conditions for quorum sensing[101] and multispecies synergistic interactions.[104] For example, using a microfluidic flow 0gradient device, it was shown that the competition between bacteria in mixed species biofilms composed of *Pseudomonas aeruginosa* and *Flavobacterium sp.* was highly dependent on the local flow rates (Figure 4b).[106]

Drawbacks to CLSM include cost, portability and potential for background signals due to autofluorescence from some microfluidic device materials under laser irradiation. In addition, competition between the high resolution and low working distance in most objective lenses can limit the depth of observation, resulting in certain regions of the biofilm becoming inaccessible unless the spatial resolution is reduced. Additionally, attenuation of laser excitation or fluorescence emission within the biofilm interior can cause artificially low intensities relative to the local number of fluorescent species. Therefore, photon intensity is usually treated in a qualitative manner, despite the potential for quantitative information. Multiphoton systems operating with longer-wavelength lasers can reduce this problem but at an increased cost. Another drawback is difficulty in observing the EPS without the use of specialised staining methods,[106] which can affect metabolic activity. This drawback has been addressed by confocal reflection microscopy, which uses the same optical setup as CLSM but does not require the presence of fluorophores for detection. This method has been demonstrated for biofilm studies in microchannels that generated high-quality topographical images of biofilms at during growth (Figure 4c).[107] Finally, microscopy generally cannot perform chemical characterisation directly on the modified biofilm. Solutions to this missing element are discussed in the following sections on *in situ* electrochemical, spectroscopic and other emerging analytical techniques.

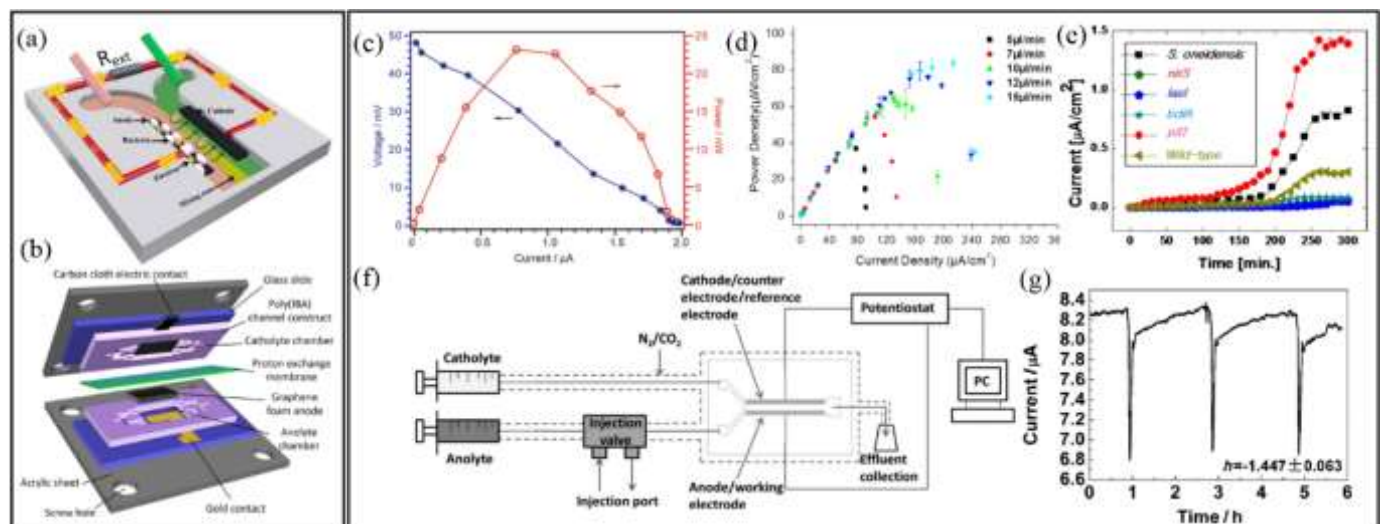

Figure 5. Two-electrode electrochemical microflow cells including (a) a membraneless microfluidic microbial fuel cell, with external resistor, $R_{ext}$, connecting the anode and cathode[117] and (b) a membrane-based microfluidic microbial fuel cell. (c) Polarization curve (blue) and power output curve (red) for a microfluidic MFC.[119] (d) Power density curves for a microfluidic MFC at different flow rates. [122] (e) Current density versus time growth curves of six different electroactive biofilms during a parallel growth experiment.[126] (f) Two-electrode bioelectrochemical device with special injection valve used to record the effect of different chemical compounds on the output current from an anode-adhered *Geobacter* biofilm. (g) Perturbations to current output from device shown in (f) during 2 min applications of sodium fumarate.[50]

## Electrochemical measurements

Electrode integration remains at the heart of the movement towards highly integrated, system-level microfluidic platforms, with examples in biosensing, point-of-care diagnostics, sample manipulation, high current applications, imaging, and other areas.[108-112] These technical successes are attributed to the diverse backgrounds of microfluidic community researchers, and a wide range of suitable materials and microfabrication techniques for creation and integration of miniaturised electrodes.[113-115] Therefore, the stage has been set for innovative electrochemical studies of biofilms in microchannels. In general, these studies focus on electroactive biofilms, which are targeted as functional materials intended for used in bioelectrochemical systems (BES) for sustainable energy, bioremediation and also for new "living sensors". In addition, certain electrochemical techniques such as electrochemical impedance spectroscopy (EIS) are applicable to non-faradaic biofilms. In the following sections, we discuss two- and three-electrode microfluidic biofilm growth cells separately because of the differences in the applicable electrochemical techniques for each case.

### Two-electrode systems

Typical two-electrode BES include microbial fuel cells (MFC), which produce usable electric current via the oxidation of dissolved organic molecules. This process is accomplished by an anode-adhered electroactive biofilm in which extracellular electron transfer is enhanced, thereby efficiently completing the Krebs cycle. Typical electroactive biofilms include those from *Geobacter* and *Shewanella* genera. MFCs are considered a sustainable energy technology because bioremediation is achieved while energy is produced.

MFCs were first reported over 90 years ago[116] and are currently one of the most significant biofilm applications in microfluidics. Laminar flow isolation of anolyte and catholyte solutions is possible in microfluidic MFCs (Figure 5a). This reduces the internal resistance of the device and decreases the cost and maintenance compared to microfluidic MFCs that use proton exchange membranes (Figure 5b).[117,118] Characterisation of two-electrode microfluidic MFCs is easily accomplished by measuring the voltage across an external resister ($R_{ext}$) between the anode and cathode (e.g., Figure 5a). The voltage measurements can be converted to current using the familiar Ohm's law, $V=IR_{ext}$. The polarization curve shows the voltage drop vs. current for different $R_{ext}$. The power output curve vs. current are calculated from the same data used to generate polarization curves with $P=V^2/R_{ext}$. Figure 5c shows typical power and polarization curves plotted in the same plot from a 1.5 µL MFC containing *Shewanella oneidensis* strain MR-1.[119] This single figure can reveal the maximum power output plus information on internal resistances, including quantification of different modes of energy losses. Such measurements have been used to demonstrate flow-induced performance improvements in microfluidic MFCs (Figure 5d).[120-124] Because the biofilms are constantly in contact with fresh nutrient media, constant growth conditions are maintained, unlike in batch reactors where continuous consumption of nutrients results in varying outputs and requires periodic replacement of the liquid phase.

Microfluidic MFCs can also been used as analytical platforms in which outputs from electroactive biofilms are screened for electrochemical activity.[125,126] For example, in one study, the small scale was exploited to parallelise multiple microchannels and enabled rapid assays of electrochemical activity from six different electroactive biofilms (Figure 5e).[126] This design increased throughput and also eliminated problems that can arise in sequential experiments due to batch-to-batch differences in nutrient solutions, temperature profiles, pumping inconsistencies, etc. In another class of applications, mature electroactive biofilms exposed to different chemical or biological analytes can convert the MFCs into living biosensors. In addition to high-throughput screening, microfluidics promotes rapid response times and excellent control over concentrations and analyte application times. Therefore, overexposure of the biofilm to certain analytes can be avoided, preventing changes in baseline outputs under control conditions, thereby improving accuracy.[118,127] For example, using a two-electrode microfluidic device with a special injection port on the anode side (Figure 5f), real-time quantitative analysis of the electrochemical activity of a *Geobacter sulfurreducens* biofilm was performed during application of various chemical stimuli and toxins.[50a] Accurate pulsation in current was induced by a periodic 2 min exposure to sodium fumarate. The short and accurate exposure time led to complete reversibility after approximately 3 h (Figure 5g). Recently a useful advancement in microfluidic MFCs was reported whereby PDMS devices were protected with parylene C coatings. This reduced oxygen permeability, supporting bench-top experiments involving anaerobic electroactive biofilms such as *Geobacter*.[50b]

**Three-electrode systems**

In three-electrode systems, the current flowing between the working and counter electrodes is usually measured while the working electrode potential is set relative to a reference electrode. This approach increases the number of applicable electrochemical techniques, using application of static potential (e.g., chronoamperometry, CA), swept potential (e.g., cyclic voltammetry, CV) and variable frequency AC voltages (e.g., EIS). These systems offer the potential for sophisticated investigation of charge transfer and reaction kinetics as well as isolation and identification of the different electrical resistances present in the system. In microfluidic channels, a precise laminar flow is directed across electrode-adhered biofilms to control mass transport. Thus microfluidic three-electrode devices offer a cost-effective alternative to hydrodynamic voltammetry with conventional rotating disk electrodes (RDE)[128-130] while reducing the experimental footprint and improving control over the liquid flow against the biofilm. It is important to acknowledge that accurate results require integrating a stable reference electrode into microfluidic channels. A recent publication shows a microfluidic three-electrode electrochemical cell with an Ag/AgCl reference electrode in a three-electrode configuration.[73] A Y-junction was used to produce side-by-side co-flowing streams, enabling isolation of the analyte stream against the working and counter electrodes and a KCl stream in contact with the calibrated reference electrode (Figure 6a). The co-flow setup facilitated constant application of the electrolyte solution against the reference electrode while enabling ion exchange with the analyte solution without requiring a complex setup. Using biofilms from *Geobacter sulfurreducens* bacteria as a transducer, a linear current response to concentration of an electron acceptor produced a calibration curve. Application of a known concentration of electron acceptor molecules was also used to monitor the short- and long-term system response to addition of solutions containing toxic molecules with good repeatability (Figure 6b). The same device enabled *in situ* measurements via cyclic voltammetry, showing increases in limiting current and cytochrome c signals during the initial growth stages (Figures 6c and d, respectively).

A different three-electrode microfluidic setup used a sequential electrode arrangement, with a gold pseudo reference electrode placed the furthest upstream (Figure 6e).[100,131a] The constant flow of fresh nutrient solution ensured that constant physiochemical conditions, in particular pH, were applied to the reference electrode, resulting in a stable reference potential throughout the experiment. By calibrating the pseudo reference potential to the potential of a known reference electrode, the results can be accurately compared with other results in the literature. This simplified the setup, avoiding the need for complex device design or a co-flowing electrolyte streams. This design has been used to monitor the effect of flow conditions of a typical nutrient solution on the de-acidification of a *Geobacter sulfurreducens* biofilm (Figure 6f).[131a] Electrochemical impedance spectroscopy is another powerful electrochemical technique (typically used in a three-electrode configuration) that can be applied to both electroactive and non-electroactive biofilms. In this approach, a low amplitude AC voltage perturbation is applied to an electrode-adhered biofilm, and the resulting AC current is recorded at different frequencies. The corresponding Nyquist or Bode plots can be modelled as an equivalent electrical circuit composed of electrical elements such as resistors, capacitors, etc.[131b] The electrical properties of biofilms subjected to the applied AC voltage at different frequencies can be evaluated in terms of the individual electrical elements in the equivalent model circuit, which correspond to real biophysical elements in the biofilm. Using a nearly identical microfluidic three-electrode flow cell as shown in Figure 6e, a microfluidic EIS study was conducted to generate Nyquist plots (Figure 6g) which capacitance and resistance were monitored to reveal subtle changes in the biofilm biomass and structure at the electrode-biofilm interface, which were not apparent from parallel microscope measurements.[100] Figure 6h shows changes to biofilm resistance with flow rate, which were interpreted as resulting from structural changes in the biofilm at electrode surface.

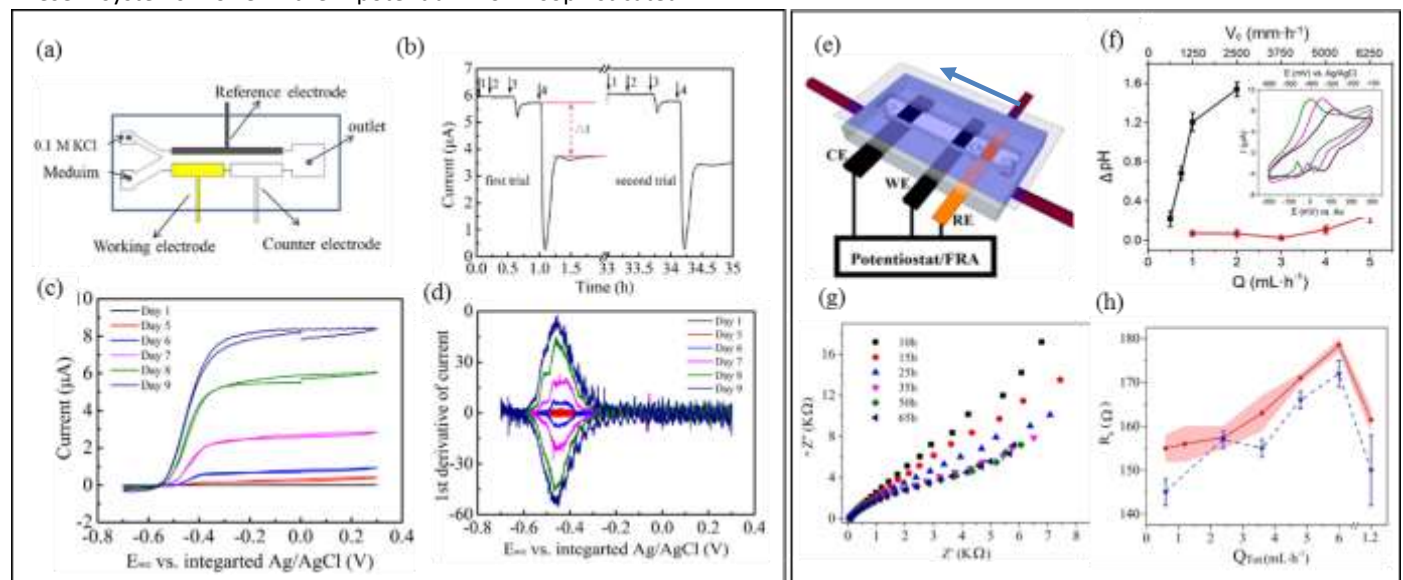

Figure 6. (a) Schematic of a three-electrode microfluidic device with co-flow configuration and (b) response from a working electrode-adhered *Geobacter sulfurreducens* biofilm showing the effect and recovery from different concentrations of a toxic compound. (c) Cyclic voltammetry results for the same *Geobacter* biofilm. (d) First derivative results from (c) showing the redox potential for the *Geobacter* cytochrome c groups over 9 days.[73] (e) Schematic of a three-electrode microfluidic device with sequential electrode arrangement and the upstream placement of a Au pseudo reference electrode. Blue arrow indicates direction of flow. (f) Results from an electrochemical study of de-acidification of *Geobacter sulfurreducens* under flow of a standard acetate nutrient solution under turnover (red) and nutrient limited concentrations (black). Inset shows the shifting CV curves that were used to monitor flow-dependant changes to biofilm pH.[131a] (g) Nyquist plots obtained from EIS in the first 65 hours of growth using similar flow cell shown in (e). Changes to biofilm resistance (h) with total flow rate ($Q_{Tot}$) modulation before (red bands) and after (blue dash lines) shear removal of biofilm upper layers.[100]

## Spectroscopy and chemical imaging

On-chip spectroscopic studies of biofilms continue to be a niche area. However, we believe that this area will be the next to experience rapid growth. Spectral microscopy can be used to attain a global understanding of structure-function relationships by performing both morphological and chemical characterisations simultaneously.[14] The importance of spectroscopic tools for biofilms is already recognised due to their ability to monitor chemical signatures on native components, including the EPS, cells, and their metabolites without significant sample preparation of addition of foreign probe molecules.[130,131] In this section, we focus on vibrational spectroscopy because technical advances in it integration with microfluidic systems is maturing, already leading to a number of relevant biofilm studies. With initial examples having already been accomplished at a technical level, the way is paved for implementation on biofilms.[134-136] To accelerate movement in this direction, we outline the basis of operation for each technique and highlight selected achievements and challenges in application to biofilms studied in microchannels.

In vibrational spectroscopy, energy exchange typically occurs between an electromagnetic field and different molecular vibrational modes of the analyte. Two popular approaches are infrared and Raman spectroscopy. Both of these techniques are useful in identifying biofilm-bound polysaccharides, proteins, DNA and RNA molecules, organic pigments (carotenoids), and lipids based on the position of the absorbance bands (Table 1).[13] Measurements of band intensity can quantify the amount of each analyte present. In addition, comparison of different band intensities can yield deeper information. For example, bioactivity is an important parameter that can be quantified by considering the polysaccharide-to-protein absorbance ratios or the relative intensities of CH, $CH_2$ and $CH_3$ bands.[137,138] Subtle changes in band position or width can reveal additional information on the biofilm and its environment but might be difficult to monitor in the crowded mid-infrared spectral window where most analysis occurs. Computer-aided analysis, including machine learning and principal component analysis, have been used in biofilm studies to find subtle correlations in complex data sets.[139,140] Furthermore, as noted below, some spectral sensing can investigate the attachment surface specifically.

Table 1. Band assignments for Raman and IR spectra of typical bacterial biofilms

| Raman Band (cm⁻¹) | IR Band (cm⁻¹) | Assignment* | | | | |
|---|---|---|---|---|---|---|
| | | Polysaccharide | Proteins | DNA/RNA | Carotenoids | Lipids |
| 3232, 2940 | 3650-3250 3100-2800 | O-H and N-H, Free OH | O-H and N-H, Free OH, C-H | - | - | O-H, C-H |
| 1750-1540 | 1760-1600 | C-C, C-O | Amide I | - | - | C-C, C-O |
| 1617-1600 | 1600-1585 | Aromatic ring | C-C Tyr, Trp | - | - | - |
| 1607 | 1500-1400 | - | C-C Try, Phe | - | - | - |
| 1580-1575 | 1550 | - | Amide II, Trp | G, A ring | - | - |
| 1460-1314 | 1450-1398 | $CH_2$, COO⁻, CH | $CH_2$, CH | - | - | $CH_2$, $CH_3$ |
| 1304-1200 | 1450,1250 | - | Amide III | G ring | - | C-H, $CH_2$ |
| 1160-1125, 1095-1000 899-855 | 1152–900 | C-C, C-O, C-O-C, C-O-H, CH | C-N, C-C, C-N | $PO_2^-$ | C-$CH_3$ | C-C, C-N, $PO_2^-$ |
| 750 | | - | Cytoc | T ring | - | - |

ARTICLE

sensitivity by a factor of $2.5 \times 10^4$, this technique also shielded

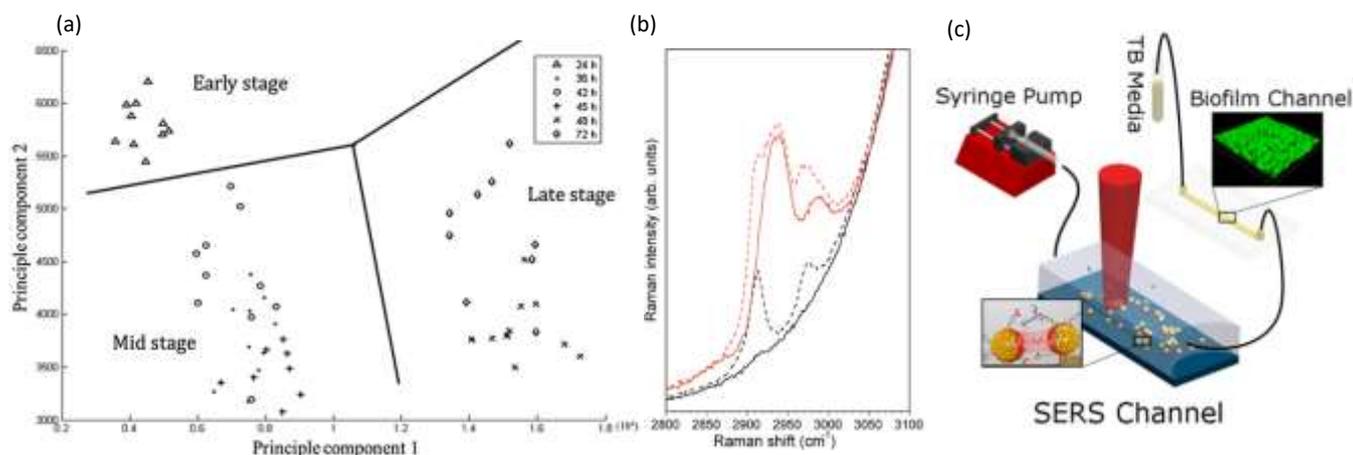

Figure 7. Raman spectroscopy in microchannels. (a) Representative two-dimensional principle component analysis of *P. aeruginosa* biofilms for identification of clustering at different developmental stages.[148] (b) Spectra of 500 mM sodium citrate solution (red) and water (black) as measured using confocal Raman spectroscopy. Solid lines showing spectra acquired in channels with surfaces coated by an opaque silver layer, thereby eliminating the background signal from the PDMS wall, and with broken lines showing comparison with spectra acquired in the non-silver coated channel.[149] (c) Schematic of the setup for measuring low-concentration bacterial metabolites in a SERS channel downstream of a growing *Pseudomonas aeruginosa* biofilm.[150]

## Raman spectroscopy

Raman spectroscopy is based on photon scattering. For molecules with a high polarizability, scattering can perturb their electron configuration, and approximately 1 in $10^6$ scattered photons undergoes a shift in energy due to induced molecular vibrations. Of note, the low polarizability of water means that water signals are largely absent from Raman spectra, which is a strong advantage over infrared (IR) spectroscopy for *in situ* measurements of hydrated biofilms without significant interferences. Another advantage is that visible-wavelength laser sources can easily access the transparent microfluidic channels for high spatial resolution studies, including spectral microscopy,[141] even at the single-cell level and in three-dimensions.[142,143]

However, additional attention should be focused on the strong autofluorescence from some microfluidic fabrication materials, specifically parylene and certain thermoplastics.[144-147] The major drawback of Raman spectroscopy is its inherently low signal. Even biofilms accumulated at the attachment surfaces are difficult to measure due to their relatively low density and small volumes compared with that of the microfluidic device material and the liquid contained within the channel. Nevertheless, device interference can be avoided with judicious selection of isolated biofilm absorbance peaks and advanced analysis techniques. For example, without any special precautions, measurements of *Pseudomonas aeruginosa* biofilms in standard PSMS/glass microchannels showed statistically significant data clusters in principal component analysis that were consistent with the early-, mid- and late-stage growth phases (Figure 7a).[148] Signal enhancement via surface enhanced Raman spectroscopy (SERS) can increase the Raman signal of the biofilm signal over the background. And as SERS is a surface sensitive technique, it is well-suited for surface adhered biofilms. A method for formation of a nanostructured SERS surface in typical PDMS/glass devices was demonstrated for early stage biofilm studies in microchannels with the use of electroless deposition followed by an air-plasma treatment to achieve nanostructuration.[149] In addition to increasing the the underlying PDMS from the probe light, thus eliminating it as a source of spectral contamination (Figure 7b). However, as illumination and detection are usually from the sample side, the biofilm can block the light from reaching the SERS surface. This problem was avoided in another study, where chemical analysis was conducted on the effluent following its interaction with a *Pseudomonas aeruginosa* biofilm using a separate downstream SERS chamber. This eliminated biofilm contact with the SERS surface and simplified both the experiment and the analysis of the results (Figure 7c).[150] Another Raman spectroscopy method involves studies of biofilms containing cytochrome c heme groups, which undergo resonance enhancement with light excitation at 532 nm, a standard laser wavelength on most Raman systems. For example, this feature has been exploited for studies of biofilms formed from the *Geobacter* genus.[150] Cytochrome c is also an indicator of bacterial phenotype[151] and can be used as a marker of cellular apoptosis.[152] Although not yet demonstrated for biofilm studies, hollow-core crystal fibres with sub-millimetre inner diameters could be used as micro-scale sensing channels for *in situ* analytical studies.[153-155]

## Infrared spectroscopy

Similar to Raman methods, Fourier transform infrared (FTIR) spectroscopy is sensitive to chemical signatures from biofilm EPS, microbes, and their metabolites.[157,158] In this method, molecular vibrations are induced via direct photon absorption by polar molecules, leading to much higher sensitivities than in Raman spectroscopy. The main drawback in applying infrared spectroscopy to hydrated biofilms is that the signals of interest can be obscured by strong water absorption in the mid-IR region. Therefore, IR spectroscopy is often applied on dehydrated samples. Further complications arise in sealed microchannels due to absorption by the device material. To solve these problems, intense synchrotron radiation (SR-FTIR) has been demonstrated for *in situ* spectroscopy on hydrated living biosystems.[159] The results were improved using an open microchannel design that eliminated background signals from the device and enabled controlled evaporation to limit water to the device and enabled controlled evaporation to limit water to the minimum possible amount while maintaining biofilm hydration (Figure 8a).[160] Another benefit was that the highly focused synchrotron beam could be rapidly rastered across the sample to produce real-time chemical maps corresponding to distribution of different EPS chemical groups (Figure 8b). Nevertheless, due to the specialised setup, SR-FTIR is not generalisable.

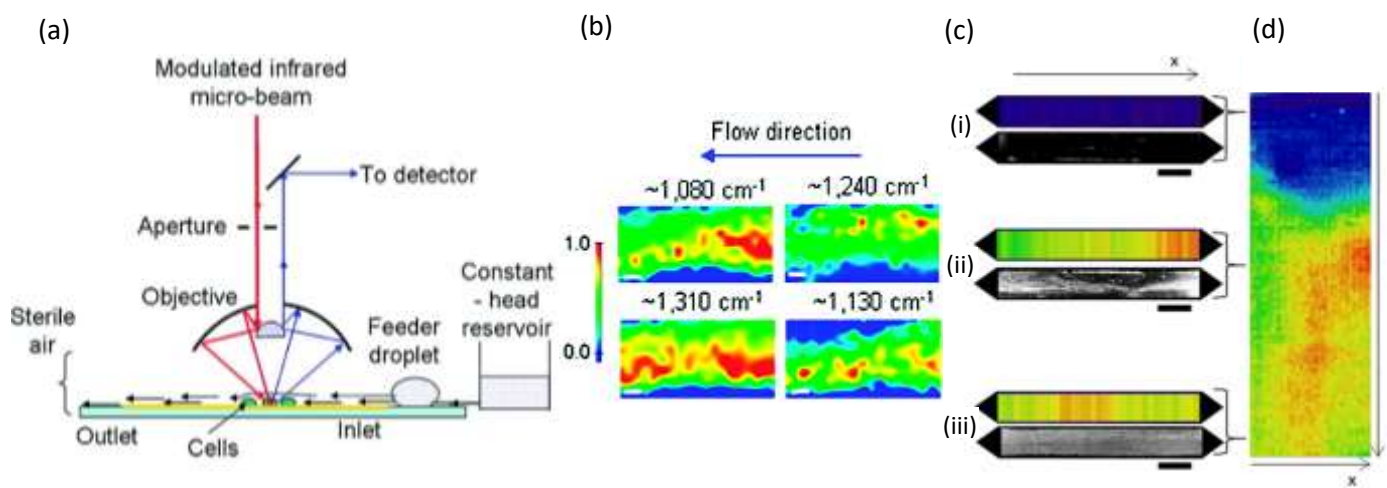

Figure 8. FTIR studies of biofilms in microchannels. (a) Schematic of synchrotron FTIR applied to *Escherichia coli* biofilms adhered to a reflective surface in an open microchannel and (b) resulting chemical maps generated from absorption bands at 1080 $cm^{-1}$ (C-O-C and C-O-P vibrations in polysaccharides and $PO_2^-$ vibrations in nucleic acids), 1130 $cm^{-1}$ (C-O vibrations in carbohydrate glycocalyx), 1240 $cm^{-1}$ (vibrations from $PO_2^-$ in DNA/RNA), and 1310 $cm^{-1}$ (amide III peak in proteins).[160] (c) Dual imaging of growing *Pseudomonas sp.* biofilm growth on top of an embedded germanium multibounce ATR element in a microfluidic device consisting of linear chemical maps from absorption band at 1540 $cm^{-1}$ (amide II signal from proteins) and dark-field optical microscopy after 5 h (i), 30 h (ii) and 60 h (iii). (d) Space-time image showing the variation of amide II absorbance along the channel length (horisontal) at different times (vertical).

Another more accessible approach is the use of attenuated total reflection (ATR), which differentiates itself as a surface-sensitive technique, achieving a path length of less than 3 μm in hydrated samples due to the limited penetration depth of the evanescent field from the surface into the sample. Integration into microchannels has been well developed following the initial reports nearly 10 years ago,[135,161] and as a surface-sensitive technique, ATR is well suited to the study of biofilms. Recently, a new approach was demonstrated in which the excitation beam was steered to different locations on a multi-bounce ATR crystal to monitor different parallel microchannels or to monitor different locations along the length of one microchannel to create one-dimensional images.[162] Parallel optical imaging on the same *Pseudomonas sp.* biofilm was achieved by coupling with a microscope in dark-field mode (Figure 8c). Continuous measurements in time generated position-time plots (Figure 8d) showing trends throughout the experiment.

## Other techniques and emerging approaches

Lastly, we review biofilm characterisation modalities that have a high probability of use in microfluidic studies, although they might have only recently been demonstrated or not yet even proven on-chip.

### Advanced microscopy:
In addition to the standard techniques discussed in the first part of this paper, new high performance imaging methods hold much potential for future biofilm studies in microflow cells. For example, studies using white light interferometry and holography have been demonstrated in on- and off-chip applications and could be easily be extended to on-chip biofilm studies. Continued development is likely due to an impressive axial resolution on the nanometre scale.[163-166] In addition, other "super resolution" microscopy techniques have been applied to biofilm research[167] and might be applied to future on-chip studies as their integration with microfluidics matures. Particle tracking methods are also promising. One method, particle image velocimetry (PIV), is a popular technique used in the study of flow profiles. Micro-PIV is often applied to verify predictions from numerical simulations in microchannels, and is especially useful for experimental measurement of flow profiles in complex flow environments (eg. microchannels containing biofilms) that are not easily modelled. Applications to simple agitated biofilm growth cells have been demonstrated,[168] but not yet for microfluidic biofilm studies, though this is likely due to its popularity in other areas of microfluidics. Another tracking method, known as micro-rheology, monitors the Brownian motion of microparticles in different target materials as a passive monitor of the local mechanical properties. It was first applied to biofilms 10 years ago with various follow-up studies since then,[169,170] including a recent on-chip study that should stimulate further work.[171,172] In that study, a parallel channel microfluidic device collected particle diffusion coefficient measurements and identification of 5 different *Sinorhizobium meliloti* bacterial strains during the first 4 days of growth. Nanoparticle tracking analysis (NTA) is a relatively new technique that monitors the motion of nanoparticles via the light scattered by the particles in dark-field mode. The smaller size of particles in NTA compared with those used in micro-rheology can improve sensitivity, but this technique is neither well tested on biofilms in regular growth cells nor on those in microchannels.

Table 2. A comparison of different tools to different analytical applications.

|  | Structural | Chemical | Mechanical | Surface | Metabolic activity |
|---|---|---|---|---|---|
| Microscopy | Yes | With additional probes | Via time-lapse and micro-rheology | Via SPRi, TIR | Via Monod kinetics during growth |
| Electrochemistry | Via ESI | Yes | No | Yes | Yes |
| Spectroscopy | Via spectral imaging | Yes | Via NMR | Via ATR-FTIR, SERS | Yes |
| Mass spectrometry | Via mapping techniques | Yes | No | Via some variants, eg. SIMS | Yes |
| Respirometry | No | Largely limited to $CO_2$ | No | No | Yes |

Strictly surface imaging techniques that have been used previously for biofilm research include surface plasmon resonance imaging (SPRi)[173] and total internal reflection (TIR) imaging.[174] Both are able to visualise sub-monolayer alterations at a sensing surface. In the case of SPRi, these are based on quantitative changes to optical density, whereas for TIR, quantitative information about cell size and distance for the surface can be obtained. While the extreme surface sensitivity can obtain high contrast images from cells and other adsorbates just a few nanometers from the surface, the lateral field of view is on the centimetre-scale, which is sufficiently large to observe long-range behaviour and heterogeneity at the same time.

**Electrochemistry:**
Electrical measurements from two-electrode devices are well-established, but with the achievement of new microfluidic three-electrode configurations discussed above, it is anticipated that hydrodynamic potentiometric biofilm studies will proliferate. As well, microfluidic devices fabricated using flexible, low-cost techniques such as material printers, printed circuit boards have enabled high electrode-density arrays and easy integration into microchannels opening the for biological applications,[175] electrochemical imaging[112] and other novel applications.[108] Future application may include scanning electrochemical microscopy (SECM), a very versatile tool for precision three-dimensional electrochemical imaging which combines the high spatial resolution of scanning probes, such as AFM with electrochemical sensing of voltammetric techniques.[97,176,177] SECM has already been demonstrated for off-chip biofilm studies,[178,179] has great potential by imaging and has the potential to be used on-chip by exploiting previous advances in on-chip AFM.[96]

**Spectroscopy:**
Among the most promising emerging spectral techniques is nuclear magnetic resonance (NMR) and magnetic resonance imaging (MRI) applied on-chip. Much interest was generated from the announcement of on-chip NMR by "remote detection" in 2005,[180,181] but we are not aware of any studies using this technique for biofilm studies. Another promising approach is MRI, which enables visualization of biofilm spatial and chemical heterogeneity via a range of *in situ* measurements, often within the same platform.[182] These measurements include high-resolution two- and three-dimensional mapping of structures, chemical groups, liquid velocities, and diffusion coefficients as well as localised measurements by spectroscopy, relaxometry, mechanical properties and porosity.[14] A recent report thoroughly demonstrated MRI capabilities on *Shewanella oneidensis* biofilms within a microfluidic flow cell, opening the door to a new area of microfluidic bioanalytical science.[183]

**Mass spectrometry:**
Mass spectrometry is growing in popularity in terms of its application to biofilm studies, identification of bacteria, biochemical groups and drug molecules.[184-187] Some examples are available in the literature demonstrating the use of mass spectrometry to study biofilms in microchannels. For example time-of-flight secondary ion mass spectrometry (ToF-SIMS) has been used for chemical imaging of fragments from different fatty acid molecules from biofilm EPS and bacteria.[188,189] This approach involved the use of a special microfluidic device which enabled controlled ion beam milling of a thin (ca. 100 nm) silicon nitride biofilm support layer, opening 2-3 μm diameter observation windows. While the technique causes the irreversible local alteration of the device and the biofilm, many such windows can be milled without compromising the overall integrity of the SiN layer, enabling acquisition of replicate measurements. It should be noted that in addition to ToF-SIMS similar self-contained microfluidic vacuum devices have been used for other for imaging modalities for aqueous surfaces in microchannels such as electron microscopy, NMR, CLSM and electrochemistry,[183,190-194] opening the way for deeper *in situ* biofilm studies in the future.

**Respirometry:**
Respirometry is another emerging approach enabling the real-time monitoring of biofilm metabolic activity via detection of molecular metabolites. In such work, a biofilm is grown on a gas permeable surface, such as the inner side of a Teflon tube, while a gas analyser collects and senses changes to $CO_2$ levels passing through to a controlled gas sheath on the other side. Researchers have used this to monitor activity of various *Pseudomonas* biofilms as a function of time,[13a] properties of the growth solution and biofilm architecture,[194] as well as in multi-species biofilms.[195] To date, these studies have been conducted in small inner diameter tubes, technically classified as millifluidic, but could be compatible with traditional microfluidic devices from PDMS, which are also known to be permeable to $CO_2$ and other small gas molecules. Another approach to respirometry has recently been demonstrated using a calorimetry, which can measure heat released by metabolically active microorganisms. This technique has been demonstrated with on-chip measurements of biofilm activity during exposure to antibiotics, with specific benefits including fast thermal equilibrium times and low medium consumption.[13b]

## Conclusions and outlooks

Bacterial biofilms are living materials with a strong propensity to change their properties as a response to environmental cues. Therefore, studying them analytically represents a great challenge. This review has advanced the argument that extracting the best, most quantitative results from powerful analytical tools and quantitative microscopy requires a parallel focus on the experimental control parameters. As such, microfluidics can be an indispensable component for modern bioanalytical platforms. In this work, we exhaustively reviewed the literature for microfluidic studies of bacterial biofilms with a focus on the historical and future trajectory of different approaches. We specifically focused on the most important on-chip characterisation techniques, including microscopy, electrochemical measurements, spectroscopy and chemical imaging. We also provided an outlook on emerging techniques that have the potential to make important impacts, including mass spectrometry and respirometry. The goal of integrating modern analytical chemistry tools in microchannels to study complex biological systems is a driver for continued technical development in microfluidics. Therefore, we can consider the state of the art in microfluidic biofilm research as a good gauge in the overall progress in microfluidic analytical platforms.

Next we provide a summary comparison between the different analytical modalities discussed in this work and their ability to perform measurements of biofilm structural, chemical, mechanical, surface and metabolic properties. Table 2 organises details from the previous sections regarding the ability to meet these requirements using microscopy (all modes conserved), electrochemistry and spectroscopy, as well as major emerging modalities, mass spectrometry and respirometry.

Structural measurements are naturally provided by microscopy techniques, which can be quantitative with computer aided image analysis. Spatially resolved spectroscopy and mass spectrometry techniques such as ToF-SIMS can deepen the spatial analysis by providing chemical structural maps. Some electrochemical techniques, namely EIS, can provide indirect information on biofilm structural properties based on electrical equivalence elements. Spectroscopy and mass spectrometry are the most naturally suited for direct measurements of a wide range of chemical groups simultaneously. Electrochemical measurements can open the way for targeted studies of redox active biochemical groups, such as cytochrome c proteins, via potential scan techniques. On the other hand, chemical analysis by respirometry and microscopy, are limited in their scope, being isolated to very specific biofilm metabolites (usually $CO_2$) or based on sensitivity of optically active probe molecules or materials (eg. pH or ionic strength). Measurements of mechanical properties are generally considered to be difficult in a microchannel due to the limitations of physically accessing the biofilm within the enclosed channel. Nevertheless, time-lapse optical microscopy combined with control over shear flow fields can provide an excellent work-around. Micro-rheology also brings a new, passive, approach to measuring a host of mechanical properties *in situ*. Surface measurements can be useful given the dominance of questions regarding biofilm attachment. These measurements can be readily conducted by integrating a sensing surface as one of the microchannel walls. For example, these include electrochemical measurements at an electrode surface, advance imaging techniques such as SPRi and TIR and spectroscopic measurements using an ATR element or SERS surfaces. Some mass spectroscopy, such as ToF-SIMS, can probe the attachment surface during its erosion by an ion beam. Finally, metabolic measurements are ideally made using respirometry techniques. However, almost every other analytical modality can provide insights as well. The versatility of spectroscopy and mass spectrometry enables the characterisation of metabolites, whose rate of production and accumulated concentrations within the biofilm are regularly used as proxies for metabolic activity. On the other hand, microscopy can usually only report on metabolic activity during initial growth via Monod kinetics.

In the next 3-5 years, we predict that general trends in lab-on-a-chip development will continue to be transferred to new biofilm research methodology. This includes the tendency towards more options for *in situ* multi-modal characterisation and high throughput analysis. Specific analytical advances for biofilm studies will likely also include a focus on chemical imaging and surface-sensitive approaches in order to match characterisation applicability to biofilm heterogeneity and surface dwelling nature, respectively. At the same time, a greater focus should be applied to parallel assaying capabilities to ensure that results are statistically significant and transferable to other studies. In the longer term, all of these advances should be integrated into a system-level configuration. This should include a streamlined user interface to achieve control over flow conditions and data acquisition, so that users can focus on scientific enquiry, without being burdened by technical details.

## Conflicts of interest

There are no conflicts to declare.

## Acknowledgements

J.G. is the recipient of an Early Researcher Award and an AUDACE grant (high risk, high reward) to study microbiological systems using microfluidics from the Fonds de recherche du Québec—nature et technologies (FRQNT). This research was supported by funding through the Natural Sciences and Engineering Research Council, Canada.

## Notes and references